\documentclass[graybox, natbib]{svmult}
\bibpunct{(}{)}{;}{a}{}{,} % suppress commas between author-names and year

\usepackage{mathptmx}       % selects Times Roman as basic font
\usepackage{helvet}         % selects Helvetica as sans-serif font
\usepackage{courier}        % selects Courier as typewriter font
\usepackage{type1cm}        % activate if the above 3 fonts are
\usepackage{makeidx}         % allows index generation
\usepackage{graphicx}        % standard LaTeX graphics tool
\usepackage{multicol}        % used for the two-column index
\usepackage[bottom]{footmisc}% places footnotes at page bottom
\usepackage[normalem]{ulem}	% for strike-through of text with \sout{}  
\usepackage[breaklinks=true,pdfborder=0 0 0,pdftex,hidelinks]{hyperref}  %for hyperlinks
\usepackage{soul}   % for high-lighting of text

\newif\ifhighlight\highlightfalse

\ifhighlight
\newcommand{\hbindex}[1]{\hl{#1}\index{#1}}  %highlights index entries
\newcommand{\hbindextwo}[2]{\hl{#1}\index{#2}}  %highlights index entries
\else
\newcommand{\hbindex}[1]{#1\index{#1}}  %highlights index entries
\newcommand{\hbindextwo}[2]{#1\index{#2}}  %highlights index entries        
\fi

\usepackage[dvipsnames]{xcolor} % color names: https://en.wikibooks.org/wiki/LaTeX/Colors

\makeindex             % used for the subject index
\usepackage{soul}

\begin{document}

\title*{Open-Science Platform for Computational Materials Science: 
AiiDA and the Materials Cloud}
\titlerunning{Open-Science Platform for Computational Materials Science} % Shorter title version
\author{Giovanni Pizzi}
\institute{Giovanni Pizzi \at Theory and Simulation of Materials (THEOS), 
and National Centre for Computational Design and Discovery of Novel 
Materials (MARVEL),  \'Ecole Polytechnique F\'ed\'erale de Lausanne, CH-1015 Lausanne, Switzerland, \email{giovanni.pizzi@epfl.ch}}

\maketitle

\vspace{-2cm}
\emph{This document has been published as part of the Handbook of Materials Modeling (Methods: Theory and Modeling), edited by W. Andreoni and S. Yip, in 2018, so it refers to the status of the described platforms at that point in time.}

\vspace{2cm}

\abstract{We discuss here our vision for an Open-Science platform for
computational Materials Science. Such a platform needs to rely on three pillars, consisting of 1) open data generation tools (including the simulation codes, the scientific workflows and the infrastructure for automation and provenance-tracking); 2) an open integration platform where these tools interact in an easily accessible way and computations are coordinated by automated workflows; and 3) support for seamless code and data sharing through
portals that are FAIR-compliant and compatible with data-management
plans. As a practical implementation, we show how such a platform in a few examples and focusing in particular on the combination of the AiiDA infrastructure and the Materials Cloud web portal.}

%\tableofcontents

\section{Introduction}
In the field of atomistic materials science, computer simulations
have become a key ingredient in materials design. The availability of 
accurate codes based on density-functional theory (DFT) and beyond-DFT methods and the ever-increasing speed of supercomputers make computational
materials science more accessible. Indeed, computations have developed into an essential component to complement experiments in the study and optimisation of materials properties.
The current relevance of these techniques is also demonstrated 
by the fact that twelve out of the top-100 most-cited papers in the
whole scientific literature are about DFT-related methods~\citep{VanNoorden:2014}.
As a consequence, many groups have started computing materials
properties for large sets of known and unknown materials, often
starting from databases of crystal structures like the ICSD~\citep{Belsky:2002}, the COD~\citep{Grazulis:2012} and the Pauling File~\citep{Villars:2004}. Many of these computed materials properties
are available online; some of these portals are also described in this handbook, and include the Materials Project\citep{Jain:2011}, AFLOWlib\citep{Curtarolo:2012}, OQMD\citep{Saal:2013}, Nomad~\citep{nomad} and the CMR\citep{Landis:2012}. 

Furthermore, large sets of computations are nowadays being used
as training data to predict materials properties more efficiently using machine-learning techniques~\citep{Ramakrishnan:2014,Dragoni:2018}.
In general, however, 
it is essential to have enough information on the data, including
how it was generated (i.e., its provenance), 
which physical and numerical parameters
were used, and in general be able to reproduce the results to validate
them. 
This is even more important for machine-learning, where
accurate predictions are possible only if the quality of the 
data used to train the algorithms is known and consistent across datapoints.

In addition to this, 
the availability of open data is beneficial to boost research and discovery because datasets can be repurposed for new studies and 
analyses not considered by the original authors.
In principle, recording data with its full provenance and 
sharing it should be much easier for computer simulations with 
respect to, e.g., experiments. However, in practice there are
a number of aspects that hinder automatic computation
of materials properties, calculation reproducibility, dissemination of data
provenance, and sharing of open research data.
In this chapter, we discuss our vision of a platform for Open Science
that can lower these barriers and we review the challenges that need to 
be addressed. We also present the two software infrastructures
\hbindex{AiiDA}~\citep{Pizzi:2016} and \hbindex{Materials Cloud}~\citep{MaterialsCloud} and show examples of how their combination 
makes it possible to create a fruitful Open Science ecosystem.

\subsection{\label{sec:pillars}The pillars of an Open-Science Platform}
Before starting, we want to define the term
\hbindex{Open-Science Platform} (OSP) and how it is used in this chapter.
With \hbindex{Open Science}, we refer to a combination of open tools and data that make 
it possible to run simulations and then share and reuse the results without
barriers, with the aim to accelerate scientific discovery.
While open data is definitely an essential ingredient of an OSP, we believe that
the platform must have a strong focus on the tools to generate
and share the data.
Moreover, it should be composed by modular components, so as to
cover a multitude of use-cases and to encourage researchers 
to use and expand it with contributions.

We think that an OSP should be based around the three following pillars: 1) open data generation tools, including open simulation codes, 
an open architecture to manage simulations, and 
open \hbindex{workflows} to steer them; 2) an integration
platform that makes these tools accessible and available in the form of automated solutions, not only to experienced computational 
researchers but also to experimentalists, students or the industry;
and 3) support for seamless data sharing through
portals that make data not only findable (e.g., via DOIs) and 
openly available, but also interoperable and reusable, encouraging 
the use of open data and code licenses.

Moreover, in our vision an OSP should also include
the availability of open libraries of curated input data, often needed 
for simulations (like crystal structures or pseudopotentials) 
and that can enable the creation of automated workflows, as well as of open learning and educational resources to ease the introduction in the field of young researchers.

%%%%%%%%%%%%%%%%%%%%%%%%%%%%%%%%%%%%%%%%%%%%%%%%%%%%%%%%%%%%%%%%%%%%%%%%%%%%%%%%%%%%%%%%%%%%%%%%%
%%%%%%%%%%%%%%%%%%%%%%%%%%%%%%%%%%%%%%%%%%%%%%%%%%%%%%%%%%%%%%%%%%%%%%%%%%%%%%%%%%%%%%%%%%%%%%%%%
%%%%%%%%%%%%%%%%%%%%%%%%%%%%%%%%%%%%%%%%%%%%%%%%%%%%%%%%%%%%%%%%%%%%%%%%%%%%%%%%%%%%%%%%%%%%%%%%%
\section{\label{sec:datageneration}Open Science Pillar 1: Open data generation}
\subsection{\label{sec:opensimulationtools}Open simulation tools}
The first requirement to be able to generate data within an OSP
is the availability of open simulation codes. 
In the field of Materials Science
(and limiting to atomistic simulations using density-functional theory only)
a number of open codes are available thanks to the developments that have 
happened in the past few decades. These include Quantum ESPRESSO~\citep{Giannozzi:2017}, SIESTA~\citep{Soler:2002}, YAMBO~\citep{Marini:2009}, 
FLEUR~\citep{Bluegel:2006}, CP2K~\citep{Hutter:2014}, ABINIT~\citep{Gonze:2016} just to mention a few. This open licensing model makes
the codes accessible to everybody. Moreover, it becomes possible to build 
simulation services on top of them that can
also be beneficial to the code themselves, as these services can become financing channels
for the code development.

There are also other codes that are widespread in the community and have
commercial licences, like for instance VASP~\citep{Kresse:1996} 
or CASTEP~\citep{Clark:2005}.
These might have benefits in term of additional features implemented, 
speed, robustness, more widespread adoption or stronger user support 
(the latter is often very valued outside academia in industrial research and 
development environments).
The challenge for an OSP becomes then to be able to integrate
also these non-open tools while abiding by their license terms. 
Models can be devised that are beneficial both for the commercial codes
and for the platform. An example could be to provide open interfaces
and plugins for the codes to enable or facilitate their integration in the OSP, 
while keeping a commercial license for the codes. 
The latter can also benefit from this model because having interfaces ready can 
facilitate the code adoption by OSP users.

\subsection{The ADES model and the implementation in AiiDA}
In recent years, many research projects used a High-Throughput Computing (HTC) approach to scan 
hundreds of thousands of different systems and identify those with 
optimal materials properties.
For this kind of projects it is unrealistic to run all 
the simulations manually, and even more to control the sequence of calculations needed to compute a given materials property. 
Tools are hence needed to help manage and 
store simulations, search through them and at the same time steer their execution
when calculation dependencies exist. Our experience showed that these 
tools can easily grow in complexity if they need to be reusable and modular.
Therefore, there is a need to collect and develop them in an organised 
architecture.

In~\citep{Pizzi:2016}, a model for such a computational science architecture
to manage calculations and workflows has been discussed, based on 
the four ADES pillars of \hbindex{Automation}, Data, Environment and 
Sharing. These are at the foundation of AiiDA, a python platform
introduced in the same paper. Here, we briefly describe the \hbindex{ADES model}
and how AiiDA implements its four pillars, and later we discuss 
why an ADES-compliant architecture is essential within an OSP.

The first ADES pillar, \emph{Automation}, involves all those software components
that aim at solving the issue of managing large numbers of HTC runs on 
supercomputers. AiiDA, in order to be to be independent of the supercomputer details, implements plugin interfaces to control the connection, transfer files and execute commands (e.g. via SSH) or to interact with job schedulers. These plugins are used by
a daemon that runs in the background and is responsible for creating new
calculation inputs and uploading them to the supercomputers, submitting new simulations and managing their lifecycle on 
a job scheduler, and retrieving and parsing results when they finish.

\begin{figure}[tbp]
    \centering\includegraphics[width=0.9\linewidth]{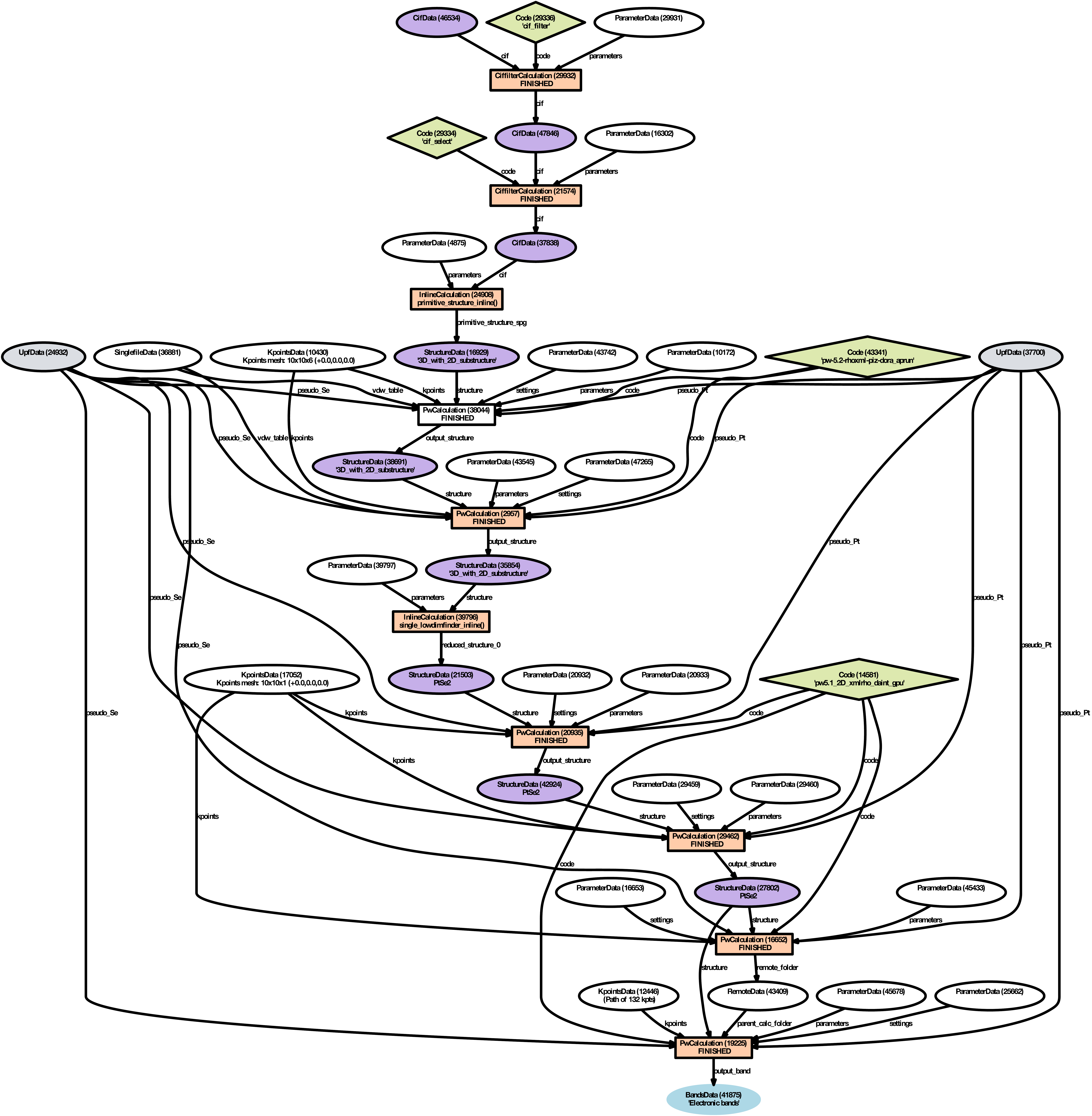}
    \caption{\label{fig:provenance} The provenance graph
    for the band structure of a 2D material (PtSe$_2$) from~\citep{Mounet:2018}.
    AiiDA automatically records the provenance and can display it in graphical form
    with the \texttt{verdi graph generate} command.
    Here, calculations are shown as orange rectangles and data as ellipses. 
    The codes used (binaries with their location, libraries) are also recorded 
    in the provenance graph (green diamonds). 
    We have highlighted with colours some specific data types: 
    pseudopotentials in grey, the final band structure in blue, and 
    crystal structures in purple. Data nodes in white are other input 
    parameters for the calculations (e.g., k-point
    grids or input keywords). The graph makes it apparent that the final
    band structure for the 2D material was obtained starting from an initial 3D structure,
    in this case from the Crystallographic Open Database (COD) (top purple node), 
    first via a set of filtering steps, then via a Quantum ESPRESSO relaxation, 
    followed by the geometrical exfoliation and a final relaxation and 
    band structure calculation for the 2D structure.}
\end{figure}

The second ADES pillar, \emph{Data}, is then needed to store and preserve all the
generated data in a reproducible and searchable way. AiiDA uses a \hbindex{provenance}
model based on \hbindex{directed acyclic graphs} to keep track of all inputs and outputs,
and of the logical relationships between different calculations. The provenance
is tracked automatically by AiiDA (see also Section~\ref{sec:ADESinOpenScience}) and
can be browsed at any time to understand how data were generated or calculations
were run. An example of a provenance graph as tracked by AiiDA is shown in Figure~\ref{fig:provenance}.
Moreover, it is also important to have the possibility to analyse
results efficiently. To this aim, AiiDA stores the \hbindex{provenance graph} 
in a database, whose schema is optimised to ensure that 
typical queries run fast.

The third ADES pillar, \emph{Environment}, focuses on two aspects. The first consists in
ensuring that AiiDA is simple to use and provides users with an effective
research environment to help and facilitate them in their work.
This happens thanks to user APIs and command-line tools that simplify the execution
of typical tasks (management of computational resources and codes, inspection of calculation inputs and results, data analytics) and by extensive documentation of these features. 
Moreover, AiiDA provides a transparent access to the database, that does not require knowledge of SQL or similar database languages, to perform queries and data analytics. This is implemented both via the use of Object-Relational Mapping (ORM) classes to access directly the data stored in the database, and via the QueryBuilder class to enable users to perform generic
queries on the AiiDA graph by using standard python syntax, independent of
the database backend.
The other defining aspect of the Environment pillar is the possibility to implement and
manage complex, nested \hbindex{scientific workflows}. These are a core ingredient
of an OSP (see also Section~\ref{sec:sharing}) and are
needed since the vast
majority of materials properties are computed by a sequence
of different calculations with non-trivial logic to control the choice of 
physical and numerical parameters, the dependencies, the convergence loops, and to perform error recovery. 
This logic is known to experienced computational scientists; encoding
it in workflows not only simplifies the management of simulations when they must be
repeated many times with different input choices, but also enables
automated computation of materials properties. The latter aspect is extremely beneficial first because these automatic workflows can be handed over to, e.g.,
experimentalists that can use them to quickly evaluate the properties they are 
interested to; but also because it becomes possible to recompute the properties
with different approximations or different codes for verification and
validation purposes, to contribute to existing databases by
new materials with their properties, and to validate the correctness of these new contributions automatically.

Finally, \emph{Sharing} focuses on creating a social ecosystem to encourage data 
and workflows reuse and accelerate research discoveries. 
AiiDA can export all data stored, including 
the whole provenance graph and not only inputs and outputs. This information
can be then imported in another AiiDA instance and used 
for new simulations or for further data analytics.
AiiDA adopts common formats for typical data structures (like
crystal structures, electronic bands, molecular-dynamics 
trajectories, \ldots) that are independent of the simulation codes. Thanks
to the automatic tracking of the provenance, moreover, these can be converted in any other standard format at any time, as we discuss with an example 
in Section~\ref{sec:aiida-tcod}. 

\subsection{\label{sec:ADESinOpenScience}ADES in Open Science}
We believe that one of the most effective approaches in the implementation
of the ADES model, especially in the context of an OSP, is to couple
Automation and Data storage. 
More specifically, AiiDA requires that all information needed to generate the 
calculation inputs is stored in the database even before the calculation is launched. 
The daemon is then responsible to create the actual input files, using only
information already present in the database.

This coupling comes at a small cost for the researchers when new calculations
are generated, as they must provide inputs using the AiiDA data structures and
run them through the daemon (even if the additional barrier can be lowered by simplifying the user interface, as prescribed by the Environment pillar). 
On the other hand, this approach ensures that common data structures
are easily reusable even in different codes and that all calculations are reproducible 
and the provenance metadata is always correct. In fact, if the user had to add provenance information only after the execution of thousands of simulations,
it would be a huge and tedious work, especially if calculations are not 
homogeneous. For this reason, in most cases
researchers end up not adding complete provenance information to their data,
unless strictly necessary. Moreover, this approach is error-prone as, e.g., 
the wrong inputs could be assigned to a given calculation.

The coupling of Automation and Data, simplified in a proper Environment,
is then strengthened (especially in view of our discussion on Open Science)
by the Sharing capabilities of data, provenance and workflows. 
In the next sections we will show with a number of examples how an ADES tool becomes an essential ingredient of an OSP and completes 
it when coupled with additional services.

\section{\label{sec:accessibleopenscience}Open Science Pillar 2: Making it accessible}
\subsection{\label{sec:sharing}Not only data: Sharing of workflows and plugins}
Sharing research data is one of the components of Open Science, as it allows other
researchers to reproduce and reuse the results, perform new data analytics and
start new research using the data published.
Hence, data sharing in an open and reusable format is essential.
To this aim, recently, in the Materials Science community a number of sharing portals have appeared, some of which are also discussed in this Handbook, like
the Materials Project, AFLOWlib, OQMD, Nomad and the CMR. 

We emphasise, however, that sharing should not only be made easy and with appropriate open licences, but also it is essential that the data is
distributed together with sufficient metadata information to 
understand it and how it was generated. 
If simulations are run with an ADES tool like AiiDA that tracks data provenance, this aspect is simply addressed by always sharing data together with the corresponding provenance. 
Furthermore, data and metadata should be shared in a standard format.
Defining standard formats is a community effort and cannot be done by 
a single research group. Nevertheless, automatic provenance tracking
makes it possible to export data after its generation in any 
other format, as discussed in Section~\ref{sec:aiida-tcod}, and hence
any future standard can also be supported with very limited effort.

However, we believe that Open Science
cannot be limited to open data sharing, as reproducibility is hindered if the
tools used to generate the data are not available.
These tools include the quantum simulation packages to compute materials properties
as discussed already in Section~\ref{sec:opensimulationtools}, 
but also the pre- and post-processing tools to analyse the data, and more generally
the scientific workflows to obtain a set of materials properties
from an initial minimal input, like a crystal structure. 
These are composed of a number of components: the
infrastructure to run the workflows (like AiiDA); the plugins to interface
AiiDA with the various codes adopted; and the logic that encodes the
scientist's knowledge on the choice of numerical and physical parameters, on how to perform the sequence of calculations and on how to deal with
potential errors or convergence issues. Only if all these components are openly
available, we can truly speak of reproducible Open Science. 
Indeed, we emphasise here that the availability of the provenance graph for a given dataset like the one shown in Figure~\ref{fig:provenance} is extremely useful, but often allows only to reproduce that single result. 
In many cases, however, the graph is not enough to understand the sophisticated workflow logic
used to select input data (like cutoffs or numerical parameters) or
to filter relevant results (e.g. in a high-throughput study).

The challenge for Open Science, encoded in the second pillar discussed in the Section~\ref{sec:pillars}, 
consists then in ensuring that all these 
components can be shared and reused with limited effort; that they can
be interoperable and used together; and that contributions to them by third-parties with extensions or improvements are encouraged. 

\subsection{\label{sec:encouraging}Encouraging contributions}
An OSP should be general and support a variety of simulation codes, data types and workflows.
Clearly, maintaining and supporting this ecosystem cannot be sustained by a single group, in particular for plugins to support specific simulation codes, because writing them requires in-depth knowledge of the code and of its typical usage patterns. Even more, this is true for workflows, that are both coupled to the codes used and to the specific research field or topic. 

As a consequence, in AiiDA we opted for a plugin interface. The main infrastructure, ``AiiDA core'', only contains the main logic 
that is independent of the codes, like dealing with external supercomputers, storing data and provenance in a database, or querying it. All the tools specific to codes, data and workflows are implemented as plugins in independent repositories. A design based on plugins is essential, but an effective
implementation can facilitate their installation and encourage contributions. For instance, in earlier versions of AiiDA, plugins had to be contributed to the code repository of AiiDA. Since version 0.9, the AiiDA plugin infrastructure has been improved: these can now be developed in independent repositories, and final users can easily choose which plugins to install. 
Once installed, AiiDA automatically detects and uses them.
The advantage is two-fold: first, it encourages researchers to contribute their own code, as they do not lose control over it but maintain full authorship and can even decide a custom licensing scheme. Most importantly, if plugins live in different repositories, their development can occur independently without having to tie and synchronise their releases (e.g., if one code just needs a bugfix while another one is in the process of a big refactoring and is not ready for a new release).

\begin{figure}[tbp]
\centering 
\includegraphics[width=0.7\linewidth]{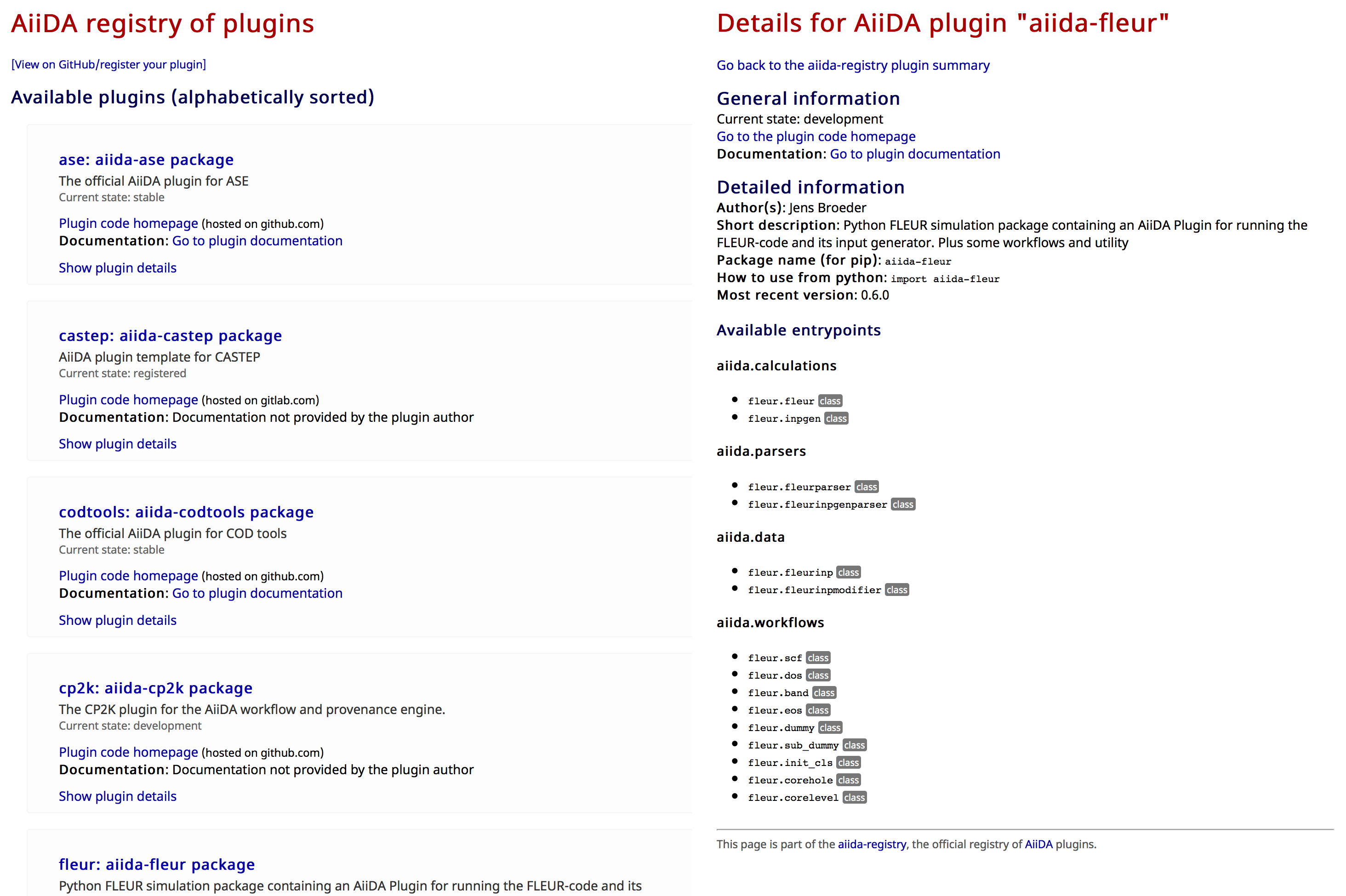}
\caption{\label{fig:pluginregistry}Left: a screenshot of 
the homepage of the AiiDA plugin registry discussed in Section~\ref{sec:encouraging} and available at \url{http://aiidateam.github.io/aiida-registry}, 
that contains a list of all available plugins
for AiiDA. Right: an example of a detail page for one of these
plugins, containing general information on which functionality 
is provided by the plugin, how to install and use it, and where
to find the code and the respective documentation.
}
\end{figure}

During the design and improvement of the plugin interface, the AiiDA team soon recognised the need of a centralised repository to list existing plugins. In fact, in the past few years it occurred that more than one researcher started to develop a plugin for the same code, driven by their needs. As a result, two very similar but essentially incompatible plugins were released. This results in work duplication with the additional risk that both plugins miss some important feature present in the other one. As a consequence, users that are faced with the choice of a plugin might get confused. Moreover, having multiple, slightly different formats also hinders sharing and reusability. 
To address this issue, the AiiDA plugin registry~\citep{AiiDARegistry} has been created (see also Figure~\ref{fig:pluginregistry}), consisting in a centralised list of existing plugins, brief notes on how to use them, and links to their code repository and documentation. The plugin code is not copied or duplicated. Instead, the registry acts merely as an index to facilitate the discovery of existing plugins.
Anybody can register a new plugin, and developers are encouraged to do so in the very early stages of development. Beside allowing them to reserve the plugin name (that needs to be unique among all plugins and should not be changed over time), this policy also reduces the risk that multiple researchers start independently to develop plugins for the same code. 
To describe the readiness of plugins, a ``state'' flag mentions if the plugin is stable and ready for production, under development or only registered. Finally, a ``plugin-template'' repository is also provided (and is also available on the plugin registry) that can be copied and modified 
to start developing a new plugin very easily.
To prove the effectiveness of this approach, we note that just one year after the creation of the registry 17 different plugins are already available, 15 of which are stable or under development, including plugins for widespread simulation codes like Quantum ESPRESSO, VASP, CP2K, FLEUR, SIESTA and YAMBO. Moreover, more than half of them provide detailed documentation websites, also in this case facilitated by the backbone documentation structure provided by the plugin template. 

Finally, also in the spirit of facilitating usage and contributions and of increasing interoperability, AiiDA has been integrated and made compatible with other libraries and tools written in python, that is becoming an extremely popular programming language in the materials science community. For instance, AiiDA can import and export structures between its internal format and the format of ASE~\citep{Larsen:2017} and pymatgen~\citep{Ong:2013}, so that functionality present in these two libraries (crystal structure manipulation, comparison, processing) can be seamlessly integrated in any AiiDA workflow. Similarly, crystal symmetry can be analysed using \hbindex{spglib}~\citep{Togo:2018}, and high-symmetry $k$-points in the Brillouin zone (together with suggested paths for band structures) can be automatically obtained via \hbindex{seekpath}~\citep{Hinuma:2017} and directly used as input to calculations.

\subsection{Documentation, tutorials and open learning resources}
The availability of open codes and tools for data generation and analysis is, as described in the previous section, an essential aspect of Open Science. If, however, these tools are not straightforward to use, their adoption is strongly hindered. We believe therefore that it is essential to invest resources in lowering the barrier for users.
To achieve this goal, it is needed to have well-written and extensive 
documentation for the codes, describing how to install them and how 
to specify their inputs. In our experience, it is important
to ensure that the documentation is easily searchable and indexed
in web engines. 
Moreover, it is very effective to provide
quick-start guides, example scripts or input files that can be copied and 
adapted by users, and a ``cookbook'' of useful recipes that are typically repeated by many users, like how to achieve a particular 
task or how to troubleshoot potential problems.
In our experience, we found the use of a 
notebook-like format particularly effective 
(in particular with the use of \hbindex{jupyter}, 
that we describe in more detail in Section~\ref{sec:matcloud} 
with some examples of its use in the Materials Cloud platform).
On top of this, documentation specifically designed for developers is greatly useful
for contributors to understand the code and be able to independently 
add new features, and is therefore another tool to encourage contributions to the platform (as discussed in Section~\ref{sec:encouraging}).

However, documentation alone (while essential) can easily become overwhelming for large codes and therefore might not lower enough the entrance barrier.
Also tutorials and schools are extremely powerful
tools for code dissemination. These events put researchers in direct 
contact with the developers of the tools, 
helping in creating a network and direct communication channels.
The advantage is on both sides: users can get direct
benefit by faster and more effective learning, and developers
can profit from these events to know how their software is used, 
to collect useful feedback on common use-cases and to
understand how to improve the code usability. 

Finally, we also believe that it is important to provide
access to educational resources like recordings of lectures, tutorials and schools that cover both the basics of the science of the field as well as the code usage and applications. These resources should be disseminated in an open format, providing access to all students and young researchers. An example of this is shown in Section~\ref{sec:matcloud}, where we discuss the Learn section of Materials Cloud, a hub for educational videos and resources in the field of Materials Science.

\subsection{\label{sec:quantummobile}Virtual machines and the Quantum Mobile}
Researchers often have a specific problem to address and look for a software
that can solve it. A good documentation and effective tutorials can help 
convince them of the functionality of the code, and jupyter notebooks can
facilitate its use, but the main initial barrier of installation and 
configuration remains.
Unfortunately, the potential issues that the users can experience
are quite varied, very dependent on their computer configuration
and are often related to code 
dependencies (need for C/fortran libraries, specific compilers or 
python packages; operating system incompatibilities; conflicts 
between multiple library versions; \ldots). 
If the installation turns out to be too complex, the researchers might be
discouraged (especially if they are not certain that the code can help 
them). The situation can be further complicated by
the need of using more than on tool at the same time and making them interact.

To address this barrier, we suggest a few different approaches to reduce and almost remove the time require to start using a software (or a collection 
of softwares).

The simplest solution that does not require to set up any online server
is the creation of virtual machines with useful software 
preinstalled. The distribution is very simple, consisting in sharing a 
single file (typically of a few GB in size) with the virtual machine image.
Once this file is obtained, by just running the virtual machine users find
all codes preinstalled and preconfigured and can directly start to run 
simulations.
A practical example of this is the 
\hbindex{Quantum Mobile}~\citep{QuantumMobile} virtual machine. 
It contains a number of simulation
codes preinstalled (including Quantum ESPRESSO, FLEUR, SIESTA, Yambo, CP2K, Wannier90) as well as AiiDA, preconfigured to run these codes within the virtual machine and coming already with a database preconfigured.
This setup is ideal for education in computational materials science classrooms,
where students can focus directly on the problem at hand and on understanding 
the results. This has been already proven in the 2018 ``Molecular and Materials 
Modelling'' course at ETH Z\"urich, in which the
Quantum Mobile has been used as the platform to run the simulations, only after
less than four months after its first release.

For larger simulations, running within the Virtual Machine is not ideal or even
possible because of CPU or memory limitations, but researchers can still use
it by configuring AiiDA to connect to the supercomputers they have access to.
An alternative virtualisation solution includes the use of Docker,
that can be thought as a tool to create lightweight virtual machines, where
the linux kernel and some resources are shared and not emulated.
AiiDA now comes with a Docker image preconfigured, and this has pushed
also many of the codes mentioned before to provide their own Docker images. 
The Docker setup provides a very similar level of
containerisation and code setup reproducibility of a virtual machine 
and is much more lightweight (for instance in terms of 
disk usage when multiple similar machines are executed). 
Therefore, it is suited for tasks where many equivalent systems
need to be automatically created. A typical use-case is in 
continuous testing platforms. On the other hand, however,
it is worth noting that the use of Docker images are less indicated for 
educational purposes with respect to virtual machines. In fact,
virtual machines are (still at the time of writing) much easier to 
start up without advanced knowledge of computer administration and of
docker and its related technologies. 
Additionally, a virtual machine provides seamless 
access to GUI applications (like text editors or file browsers), while
with Docker one would need to access the instance through the command line, 
creating a potential additional barrier for students.

\subsection{Supercomputer centres}
A second approach to facilitate users in the adoption of codes is 
to deploy them in supercomputer centres where the simulations are run.
Already now most \hbindex{High-Performance Computing} (HPC) centres have an application support team that compiles
various versions of the codes used by their users and provides
them as modules. The same could be done for tools to manage simulations like 
AiiDA. This might require, however, that these centres transition from 
a classical model of HPC providers to more general service providers
for, e.g., virtual machines, database services or 
long-term storage. A number of centres (like, in Europe, 
the Swiss CSCS, the Italian CINECA or the German JSC) are already in the
process of providing these new services and working together to federate
access to them; we expect that this trend will be followed by even more
centres in the next few years.

\section{\label{sec:opensciencesharing}Open Science Pillar 3: Seamless data sharing and preservation}

\subsection{\label{sec:dmp}Data management plans and FAIR compliance}
In the first phases of research, offline tools and codes are typically
used to perform and organise research, as the data generated can be large
and is typically confidential. However, when data must be shared, 
as described in the third pillar of an OSP described in Section~\ref{sec:pillars}, it becomes essential to employ suitably-designed web portals. 
In addition to this, currently many funding agencies require to comply with 
\hbindex{data management plans} (DMPs) for data dissemination and long-term
preservation. Ideally, data should be compatible
with the \hbindextwo{FAIR}{FAIR sharing} principles of sharing~\citep{Wilkinson:2016}, that require
data to be Findable, Accessible, Interoperable and Reusable.
Findability can be achieved only if research data is associated with 
persistent handles like DOIs to make it citable.
Free portals exists, like Zenodo~\citep{zenodo}, that assign DOIs to
datasets obtained from research projects and also guarantee long-term
preservation. However, while files generated by
the different codes can be uploaded on these services, this would still require
that another researcher has a compatible software installed in order 
to open and analyse the results in the format uploaded by the original author.
To remove this barrier, web portals need therefore to address also the other 
FAIR aspects to become effective OSPs.

In the next few sections, we show examples of how the goals of
being DMP and FAIR-compliant can be conveniently achieved 
using a combination of suitable web portals together 
with an ADES tool like AiiDA to manage simulations and 
track provenance. We discuss in particular the integration of AiiDA with 
the Theoretical Crystallographic Open Database (\hbindex{TCOD})
in Section~\ref{sec:aiida-tcod}, 
that proves how the automatic tracking of data provenance makes it possible 
to tag results with standard metadata automatically and after the simulations have run; and the Materials Cloud portal (Section~\ref{sec:matcloud}), that provides tools encompassing all steps of computational research, from learning to generating data and curating the results, to finally publishing and sharing results, while being fully compliant with the FAIR principles.

\subsection{\label{sec:optimade}Interoperability between different databases: the OPTiMaDe API}
Even if automated provenance tracking allows to store a posteriori the
metadata information in any format, as we discuss in the next section,
it is still extremely valuable that 
standards are defined. This, in fact, would allow easy interoperability 
between different existing databases. 
A community effort in the direction of database and web-portal interoperability
is happening in the \hbindex{OPTiMaDe} consortium, 
that is working towards the definition of a common API specification to be implemented by the different partners.
Many of the large databases in the community are already part of this consortium (including Materials Cloud that has the 
\texttt{\_mcloud\_} prefix assigned to it).
In the current version (0.9.5), OPTiMaDe already defines a REST API format that
makes it possible to query, with the same format, for the existence of 
crystal structures in different databases, with common filters like 
number of atoms, presence or absence of a chemical element, as
well as supporting database-specific fields. 

\subsection{\label{sec:aiida-tcod}Automatic \emph{a posteriori} metadata 
tagging: AiiDA and TCOD integration}
Ontologies and other standards for metadata tagging and sharing of research 
results in materials science are currently being discussed in the 
community, like e.g. the TCOD
dictionaries~\citep{Grazulis:2014} and the Nomad 
metadata~\citep{NomadMetadata}, 
even if there is not an established standard yet. 
For this reason, it is difficult to enforce a given ontology.
The lack of a single standard format in the community, however, is 
not a major issue if simulations are run with automatic 
provenance-tracking tools.
Indeed, tracking of provenance (if complete and automatic) 
allows users, \emph{\hbindextwo{a posteriori}{a posteriori metadata tagging}}, i.e., after all simulations have run,
to convert the provenance information in any other format 
by just implementing a converter. An example of this is shown in~\citep{Merkys:2017}, where methods and codes have been presented to 
convert the provenance as tracked by AiiDA to the format defined by the 
TCOD ontology. 
In this work the authors show in particular how Quantum ESPRESSO simulations managed via AiiDA are automatically tagged with metadata using to the TCOD ontology, with no user input required. 
Moreover, the implementation is modular and additional
plugins can be developed to support other simulation tools 
or other metadata formats.

\subsection{\label{sec:matcloud}Materials Cloud}
In Section~\ref{sec:datageneration} we have described how AiiDA addresses
the challenges of data generation in an OSP,
in Section~\ref{sec:accessibleopenscience} we have outlined how to make
Open Science accessible, and in Section~\ref{sec:dmp} we have emphasised
how web portals tailored to comply with the FAIR principles are an essential ingredient to achieve seamless sharing.

For this reason, Materials Cloud has been designed, implemented and deployed.
It is a web portal for Materials Science that is coupled to AiiDA and, in combination with it, addresses all three OSP pillars introduced in Section~\ref{sec:pillars}.
Materials Cloud is composed of different sections, that aim at assisting 
researchers during the full lifecycle of a scientific project.

The first section, \emph{Learn}, contains educational material like 
videos of schools, tutorials and lectures in the field of Materials Science, 
together with the corresponding material to assist learning (like slides and
exercises). The Learn section uses the SlideShot technology, a software platform that makes it possible to show the video feed of the speaker together with
the slides in high resolution, allowing viewers to quickly seek through the video via the slides thumbnails. This is coupled with a standard hardware setup to perform recordings and then import them into SlideShot.

The \emph{Work} section then focuses on the task of data generation.
This section addresses in particular the second pillar of Open Science
to make the simulation tools available in the form of automated solutions, removing access barriers and making them available not only
to experienced computational researchers but also to experimentalists or 
to students. Beside providing links to download various versions of 
the Quantum Mobile virtual machine (see Section~\ref{sec:quantummobile}), 
it provides access to computational tools with two different approaches,
suitable for different use-cases that we briefly discuss here.
The first approach of the Work section of Materials Cloud is to provide 
a set of \emph{online tools} to perform fast analysis of data, directly from 
the web browser, similar in the spirit
to, e.g., the many tools of the Bilbao Crystallographic Server~\citep{BCS}. This is particularly suited
for computations that can run in real time (i.e., with a running time of up to
a few seconds) and that can benefit from a graphical web interface. 
An example is seekpath~\citep{Hinuma:2017}, a tool that, given an input
structure, computes a standardised primitive unit cell according to the
standard definitions in the crystallography literature, like in the 
International Tables of Crystallography~\citep{ITB:2010}
and in~\citep{Parthe:1984}. The tool then returns also the labels of high-symmetry
points in the Brillouin zone together with a suggested path for the computation of band structures. 
The only required input, in this case, is a crystal 
structure (accepted in a number of common formats) and, optionally, a few numerical
parameters; interactivity happens mainly via to interactive 3D visualisers for the 
crystal structure and the Brillouin zone.

\begin{figure}[tbp]
    \centering\includegraphics[width=0.7\linewidth]{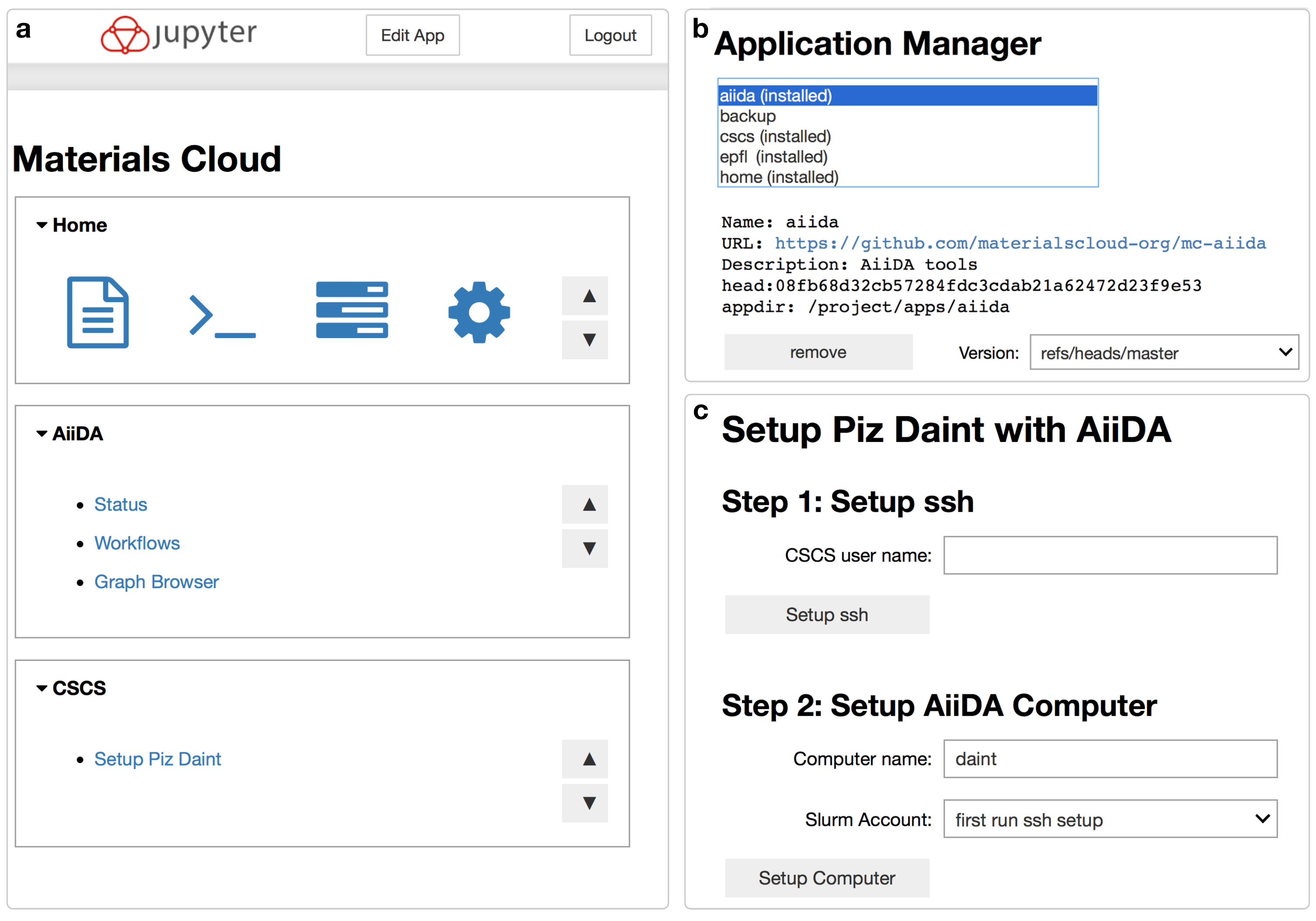}
    \caption{\label{fig:jupyter} Three screenshots of the jupyter interface
    of the Materials Cloud. Pages are customised with the AppMode plugin,
    to make the appearance very similar to a standard web page. 
    a) Home page, with the top panel with main tools to manage Materials Cloud
    applications,
    show a terminal or show the control panel. On the top, a ``Edit app''
    button allows to go back to the standard jupyter notebook interface.
    b) Application manager, that allows to select existing apps from the 
    Materials Cloud app registry and make them appear in the home page. 
    c) One of the apps to setup a new computer (in this case, the Piz
    Daint supercomputer at the Swiss Supercomputer Centre) with just
    a few clicks.}
\end{figure}

The second and more flexible approach for data generation in the Materials
Cloud is based on the \emph{jupyter} interface. Jupyter is a notebook-like 
web frontend that allows to run any python code (and also supports many other
programming languages) subdivided in cells with inputs and outputs. This
is a very flexible interface, and in Materials Cloud it is further 
powered by the AppMode plugin,
that by default hides the cells and just shows the outputs (including 
widgets like buttons, text boxes and drop-down lists) in a format that resembles a standard interactive web page. 
The jupyter interface of the Materials Cloud is shown 
Figure~\ref{fig:jupyter}, providing a home page where contributed 
apps can be added
directly via the web interface (panels a and b). Apps can be provided by anyone and just need to be registered in the Materials Cloud App registry. 
Users are provided with a working space already preconfigured with AiiDA and
codes, removing any setup time. Additional configuration (e.g., 
setting up AiiDA to interact with custom computational clusters) is significantly simplified by apps with simple GUIs 
(see Figure~\ref{fig:jupyter}c).
The only disadvantage of this section is that it requires a user login, 
because users can access a full computer with unrestricted access to code
execution and internet access, but on the other hand it is the ideal platform
to create custom fully-automated workflow solutions for the computation of 
materials properties.

Once the data has been generated (locally with AiiDA on in the jupyter section), it can be displayed in the two sections
\emph{Discover} (for curated data) and \emph{Explore} (for the ``raw'' data
as generated via AiiDA). The first section can be used by a researcher to provide an accessible interface to understand
a project and present its results with dynamic data filtering, as well as
interactive views of the figures shown in the corresponding papers.
Data can be linked to the corresponding AiiDA nodes in the Explore section. 
In the latter, data is visualised together with its provenance 
(browsable via an interactive graph to inspect the calculation 
that generated it, or to find out by which calculations it was used as input).
Calculations are always displayed together with their full set of inputs and outputs, that can also be directly viewed or downloaded. 
Materials Cloud has been designed as an extension of AiiDA so that, if the simulations are run with AiiDA, the Explore section is filled in automatically by just importing an AiiDA export file. 
To make an analogy: AiiDA is a tool used to manage simulations and provenance locally on a computer
and can be compared to git, used locally for the organisation and tracking of the history of files and source codes. Then,
the Materials Cloud Explore section plays the role of GitHub, GitLab or similar web services, providing web browsing of the 
content of repositories and acting as a central server for sharing.

\begin{figure}[tbp]
    \centering\includegraphics[width=0.9\linewidth]{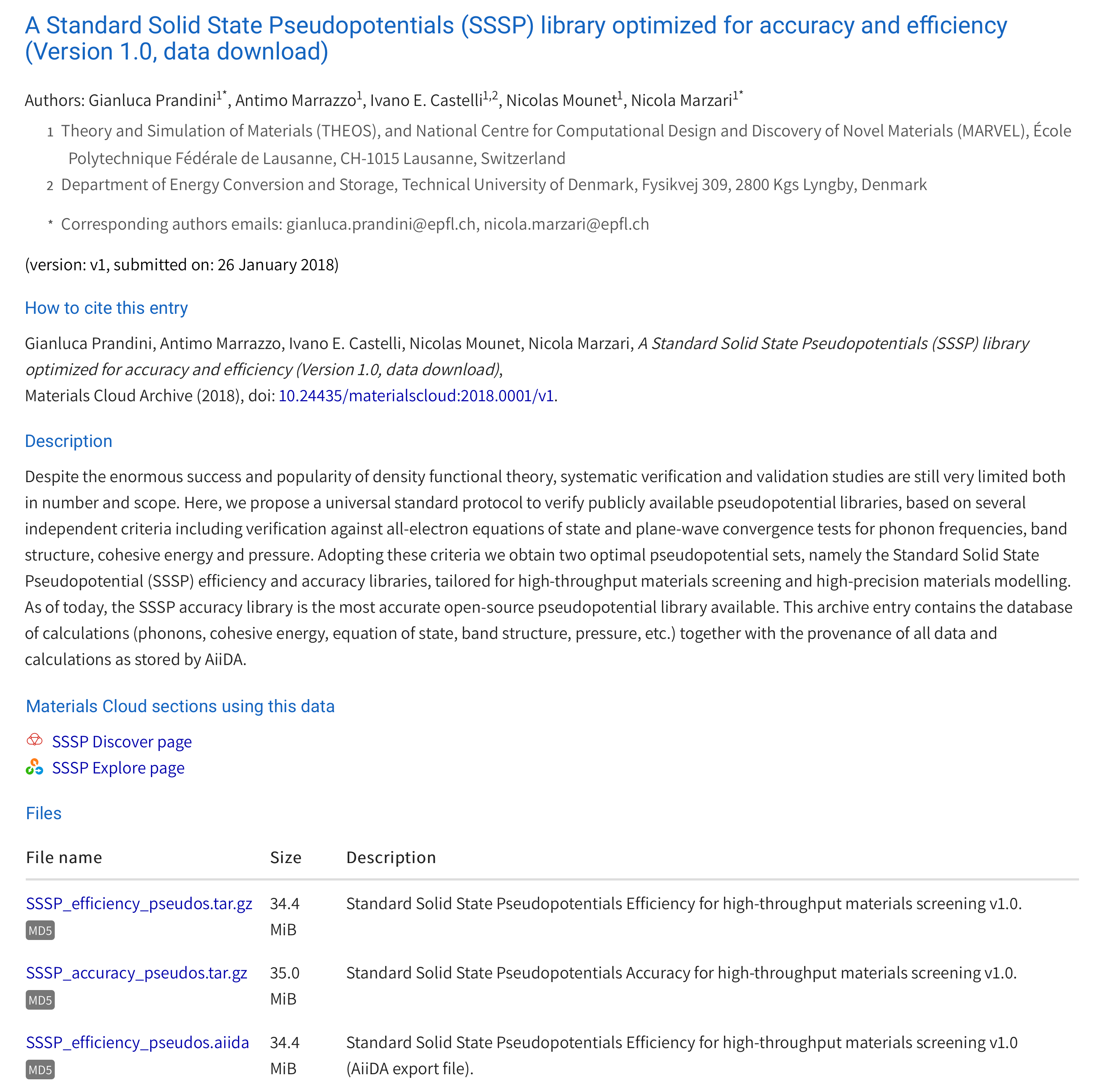}
    \caption{\label{fig:archive} An entry of the Materials Cloud Archive,
    in this case version 1 of the SSSP pseudopotential
    library~\citep{Prandini-MC:2018}. Each entry, beside typical metadata
    like authors, affiliations, and a description, can also contain a number
    of files that are preserved in the long term. Moreover, entries can have
    direct link to respective Discover or Explore sections, where data is
    directly visualisable in the web browser. 
    Materials Cloud assigns each entry a DOI 
    (the prefix of Materials Cloud is 10.24435) and requires authors to choose
    a license.}
\end{figure}

Finally, Materials Cloud has a fifth section, \emph{Archive}, for the 
long-term storage and dissemination of research results. A DOI is assigned
automatically to each submitted entry and standardised metadata are exposed
in the XML Open Archives Initiative Protocol for Metadata Harvesting
(OAI-PMH) format~\citep{xml-oai}.
Each entry can contain files with research results and can be linked to the
corresponding Discover and Explore sections.
An example screenshot of an entry of the Materials Cloud Archive is shown
in Figure~\ref{fig:archive}.

By coupling the different sections, Materials Cloud realises a
FAIR-compliant platform as discussed in Section~\ref{sec:dmp}.
In fact, the DOIs assigned to published research data in the Archive section achieve findability. At the same time, if the data has been
generated with AiiDA, each entry can be linked to a curated Discover 
section and to an Explore section, making it possible to visualise 
inputs and outputs of calculations and more generally to explore 
the data provenance.
This real-time web interface makes data accessible (second FAIR pillar)
and does not create any barrier due to authentication or to software installation. 
Interoperability (third FAIR pillar) is achieved since data
that is common to different codes (e.g., crystal structures, electronic and phonon
bands or $k-$point sets) are stored in a code-independent format and 
the outputs of a calculation can be used as the inputs for a different code.
Finally, reusability (fourth FAIR pillar) is guaranteed by the 
possibility of getting all inputs and outputs of calculations and 
reproduce them, coupled with the large selection of licenses that the users
can choose (with encouragement for those allowing for reuse, like the
Creative Commons ones). 

\section{Examples of Open-Science research using AiiDA and the Materials Cloud}
To better explain the ideas and tools described in the previous sections, 
we briefly discuss here two scientific projects, managed with AiiDA
and hosted on the Materials Cloud, that demonstrate the concepts of 
an OSP discussed here.

\subsection{The SSSP pseudopotential library}
The Standard Solid State Pseudopotentials (\hbindextwo{SSSP}{SSSP pseudopotential library})~\citep{Prandini-MC:2018}
library provides a curated selection of pseudopotentials for plane-wave 
density-functional theory (\hbindex{DFT}) codes (in UPF format).
SSSP is composed of two sublibraries, optimised respectively for \emph{accuracy}
and \emph{efficiency} by means of a number of convergence studies on elemental solids,
for various relevant physical properties including zone-boundary phonons, 
cohesive energy, pressure and band structure. 
The \hbindex{pseudopotential library} is available on the Materials Cloud Archive~\citep{Prandini-MC:2018}, see 
also Figure~\ref{fig:archive}.

In the respective Discover section, a periodic table 
uses a colour legend to indicate the optimal pseudopotential for each
element and indicates the suggested cutoff values. 
Clicking on an element shows a detailed view containing all convergence
studies and plots (equations of states, band structures, bands chessboards)
for all the pseudopotentials compared in the study. Datapoints in the interactive 
plots are clickable and bring the user to the Explore section, with browsable 
provenance for all data. Simulations were run using Quantum 
ESPRESSO and managed by AiiDA, that also tracked the provenance.

\subsection{The exfoliable 2D materials database}

\begin{figure}[tbp]
    \centering\includegraphics[width=0.9\linewidth]{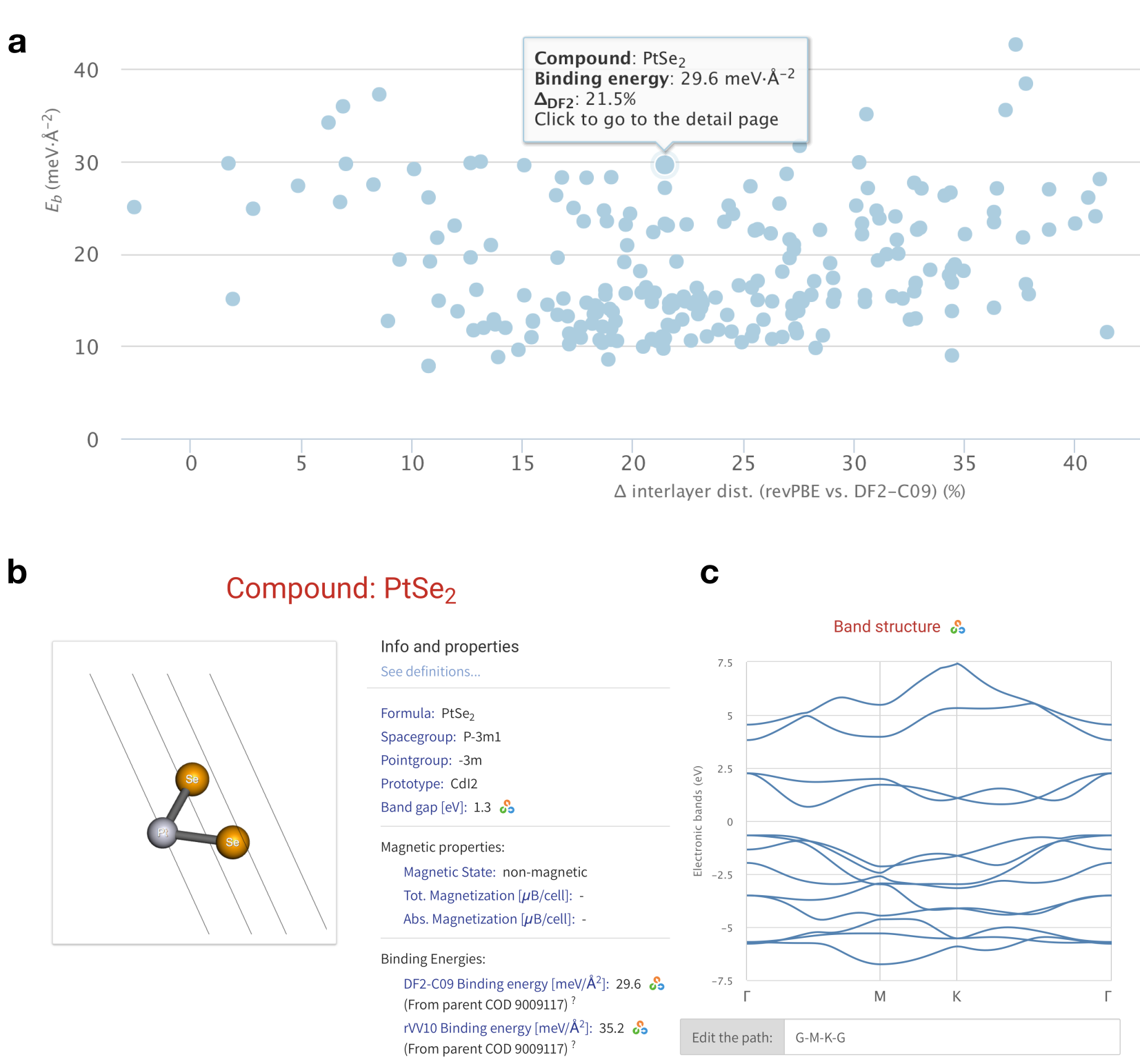}
    \caption{\label{fig:fig2d} Screenshots of the Discover section for the 
    exfoliable 2D materials section. (a) One of the three selection pages, allowing
    to select a 2D material based on its binding energy and the change
    in interlayer distance in the bulk when computed with a DFT
    energy functional including or not including van der Waals force contributions.
    Other selection views include a periodic table and a detailed table.
    (b) Top part of the detail view for one of the materials (in this case, PtSe$_2$).
    The view includes an interactive structure 3D visualiser, as well as the
    main properties computed for this material.
    (c) Another portion of the detail view showing the electronic band 
    structure of PtSe$_2$. The plot is interactive, zoomable and the default
    path (here $\Gamma-$M$-$K$-\Gamma$) can be changed by the user. Both here
    and in panel b, the small AiiDA icons are links that bring the user to 
    the corresponding Explore section, to browse the provenance of the 
    corresponding data (the provenance of the band structure is the one
    shown in Figure~\ref{fig:provenance})}
\end{figure}
    
In \citep{Mounet:2018}, the authors performed a high-throughput search of novel
two-dimensional (2D) materials by screening the ICSD and the COD databases. Starting from over 100,000 unique
bulk experimentally-known materials, they could first detect with a geometrical algorithm the $\sim$6000 that are composed by layers, then refine this list with DFT binding-energy calculations
(using Quantum ESPRESSO and managed by AiiDA) to only include those held together by weak forces. With
this approach they could identify a set of about 1800 potentially or easily
\hbindextwo{exfoliable}{exfoliable 2D materials} \hbindex{2D materials}. Furthermore, a subset of 254 materials 
(easily-exfoliable with up to 6 atoms per unit cell) has been extensively
studied to compute relevant electronic, vibrational and magnetic properties.
These results and calculations are available on the 
Materials Cloud Archive~\citep{Mounet-MC:2017}. Also in this case, the Archive section is
coupled to a curated Discover page (Figure~\ref{fig:fig2d}). In the filtering page, a 2D structure
can be selected via a table, by selecting elements in a periodic 
table or by picking a material from a binding-energy plot.
After having selected the material, a detailed view is presented, showing data
and results that include the binding energy of the 2D layer, the
interlayer distance computed with different van der Waals functionals,
the magnetic ground state (also including antiferromagnetism in larger supercells), 
the electronic band structure and the phonons. These were 
computed with the correct 2D physics that properly considers 
electrostatic screening in low-dimensional 
systems, reproducing correctly the behaviour of longitudinal and 
transverse optical phonons near $\Gamma$~\citep{Sohier:2017}.
Final results (plot and numerical values) are accompanied by small AiiDA icons,
as shown in Figure~\ref{fig:fig2d}b and \ref{fig:fig2d}c.
These are links to the corresponding Explore page, where users can
check how the results have been computed by browsing their AiiDA provenance.

\section{Conclusions}
In this chapter we have introduced our vision for an Open-Science
Platform. Such a platform should rest on three main pillars, namely: 
open data generation tools, open integration platform, and seamless
data sharing.
We have discussed the challenges set up by these requirements, and we have shown how a combination of the AiiDA code and the Materials Cloud
platform can achieve the goals defined by these three pillars.
These aspects have furthermore been demonstrated with two examples of projects that used the two tools, namely the SSSP pseudopotential library and the database of exfoliable 2D materials.
Indeed, if simulations are run with AiiDA, the generated data is
reproducible and its provenance (i.e., how it was generated)
is automatically tracked and stored. AiiDA also helps creating and 
steering automatic workflows for the calculation of materials properties.
Combined with the Materials Cloud Work section (and in particular
with the jupyter subsection), advanced workflows can be exposed with an intuitive and easy interface, minimising the barrier to access and use them. 
The other sections of Materials Cloud, furthermore,
enable sharing of computed data in a FAIR-compliant format. DOIs 
are assigned to data entries (that are preserved in the
long term) to make them findable, as required by data management plans. These entries, moreover,
are made accessible by a tight coupling with interactive views to present
curated data. The latter is linked to browsable graphs to access and explore the data provenance automatically tracked by AiiDA as well as the raw input and output files. 
AiiDA and Materials Cloud, therefore, implement our open-science vision with the aim of making science accessible to everybody, and of encouraging reuse of results to promote and support scientific discovery.

\begin{acknowledgement}
The author acknowledges the support of the NCCR MARVEL, funded by the 
Swiss National Science Foundation, and of the EU Centre of Excellence 
MaX ``MAterials design at the eXascale'' (grant no.~676598).
Moreover, the author acknowledges the work of all colleagues involved in the
design and development of the AiiDA software and the Materials Cloud platform, 
who have made the existence of these two tools possible. Alphabetically: 
Marco Borelli, Jocelyn Boullier, Andrea Cepellotti, Fernando Gargiulo, Dominik Gresch,
Rico H\"auselmann, Eric Hontz, Sebastiaan P. Huber, Boris Kozinsky, Snehal P. Kumbhar, 
Leonid Kahle, Nicola Marzari, Andrius Merkys, Nicolas Mounet, Elsa Passaro, Riccardo Sabatini,
Thomas Schulthess, Ole Sch\"utt, Leopold Talirz, Martin Uhrin, Joost VandeVondele, 
Aliaksandr Yakutovich, Spyros Zoupanos; as well as all the contributors to the 
platform in the form of suggestions, improvements or plugins.
\end{acknowledgement}

%% Shows the indexed words
%\printindex

\end{document}